%% file: main.tex
\documentclass[12pt]{article}
\usepackage{sbc-template}
\usepackage{graphicx,url}
\usepackage[brazil]{babel}   
\usepackage[utf8]{inputenc}
\usepackage[T1]{fontenc}

\usepackage{subfigure}
\usepackage{booktabs} 
\usepackage{algpseudocode}
\usepackage[portuguese,linesnumbered,ruled,vlined]{algorithm2e}

\title{Evacuação de Dados em Nuvens Ópticas com base no SLA\\ sob Cenário de Desastre}

\author{Francisco Renato C. Ara\'{u}jo\inst{1}}

\address{Programa de Pós-Graduação em Ciência da Computação (PGCOMP)\\
Instituto de Matemática -- Universidade Federal da Bahia (UFBA)\\
Salvador -- BA -- Brasil
\email{franciscorca@ufba.br}
}

\begin{document} 

\maketitle

\begin{abstract}
The popularization of cloud computing has provided the emergence of large volumes of data that are stored in Data Centers (DCs). These locations store data of different types, origins, and priorities for their owners. The DCs are subject to natural or man-made attacks. The attacks are diverse and happen quickly after their detection. Therefore, this paper proposes two techniques to evacuate data from threatened DCs to those who are outside the risk zone of the attack. The first technique is based on the Service Level Agreement (SLA) of the data and the second one is based on the order that they arrive at the DC, using the algorithm LIFO. Both techniques performed similarly on the amount of evacuated data and at the time for evacuation. However, the SLA policy distributes the data on a priority scale according to the SLA, while the LIFO policy ranks data on the same scale's priority.
\end{abstract}
     
\begin{resumo} 
A popularização da computação em nuvem tem proporcionado o surgimento de grandes volumes de dados que são armazenadas em Data Centers (DCs). Esses locais concentram dados de diversos tipos, origens e prioridades para seus proprietários. Os DCs estão sujeitos a ataques naturais ou causados pelo homem. Os ataques são diversos e acontecem rapidamente após a sua detecção. Diante disso, este trabalho propõe duas técnicas para evacuar dados de DCs ameaçados para aqueles que encontram-se fora da zona de risco do ataque. A primeira técnica é baseada no acordo de nível de serviço (SLA) dos dados e a segunda baseia-se na ordem que os mesmos chegam ao DC, usando o algoritmo LIFO. As duas técnicas apresentaram desempenho semelhantes na quantidade de dados evacuados e no tempo para a evacuação. No entanto, a política SLA distribui os dados em ma escala de prioridades de acordo com o SLA, enquanto a política LIFO classifica os dados em uma mesma prioridade da escala.
\end{resumo}

\input{1-introducao}

\input{2-trab-relacionados}

\input{4-proposta}

\input{5-experimentos}

\input{6-conclusao-e-trab-fut}


\section*{Agradecimentos}
O autor agradece a Fundação de Amparo à Pesquisa do Estado da Bahia -- FAPESB, pelo apoio financeiro a esta pesquisa.

\bibliographystyle{sbc}
\bibliography{main.bbl}

\end{document}

%% file: 1-introducao.tex
\section{Introdução} \label{sec:introducao}
Com a popularização da computação em nuvem nos variados setores da sociedade, surgiu uma crescente demanda por serviços de computação que, consequentemente, geram uma grande quantidade de dados de fontes e tipos variados que, por sua vez, são armazenados em centros de dados (do inglês, \textit{Data Centers} (DCs)). Esses locais hospedam grande parte dos conteúdos produzidos pelas aplicações e serviços existentes. Os DCs são interligados por redes ópticas de alta velocidade em que se trafega grandes volumes de dados na ordem de \textit{terabytes} a \textit{petabytes} \cite{Ferdousi2015} \cite{Li2016}.

Nos ambientes de computação em nuvem, tanto os DCs quanto a rede estão sujeitos a desastres de grandes proporções que ameaçam os dados neles armazenados, trafegados, respectivamente \cite{YuanYang2016} \cite{XiaolongXie2016}. Esses desastres podem ser naturais, tais como: terremotos, tornados, inundações, etc., e ataques causados pelo homem, como ataques nucleares, químicos, biológicos, além de armas de destruição em massa, que representam uma grande ameaça para as redes e DCs, uma vez que causam falhas correlacionadas e em cascata. Tais falhas podem causar enormes quantidades de perdas de dados e até interrupções de serviços \cite{Ferdousi2013} \cite{Xu2016}. 

Dependendo do cenário e do tipo do desastre o intervalo para sua ocorrência pode se dar em milissegundos a poucos minutos, após a detecção da ameaça \cite{Ferdousi2015} \cite{Li2016}. Isso representa um problema importante e desafiador para os serviços em nuvem entregues por redes de DC \cite{Ferdousi2014}. Tendo em vista que desastres são imprevisíveis e recorrentes, ambas, academia e indústria mantêm um esforço contínuo no estudo, desenvolvimento e aperfeiçoamento de técnicas que visam assegurar a integridade e disponibilidade dos dados no caso de eventos inesperados \cite{YuanYang2016}. 

É possível encontrar na literatura diversos métodos e técnicas que objetivam garantir a segurança dos dados, em sua grande maioria destacam-se as técnicas de redundância e evacuação de dados. A técnica de redundância é intrínseca a computação em nuvem, na qual os dados são regularmente replicados para outros DCs espalhados geograficamente pelo globo \cite{Ferdousi2015dikbiyik}. Assim, no caso de ataque a um determinado DC os dados afetados estarão disponíveis em outros locais. No entanto, essa técnica apresenta uma falha grave referente ao intervalo em que os dados são replicados, pois a mesma não está preparada para o caso de um ataque e, portanto, os dados copiados anteriormente para outros DCs podem estar desatualizados no momento do ataque, causando a perda das versões mais recentes desses dados no DC sob ataque \cite{Ferdousi2013}.

Os dados desatualizados podem causar danos irreparáveis como, por exemplo, perda de transações bancárias ou informações sobre ações de mercado \cite{Ferdousi2015}. Além disso, a perda de informações pode gerar grandes perdas financeiras e, até mesmo, perda de prestígio para as empresas provedoras de serviços em nuvem \cite{Karakoc2016}.

Já a técnica de evacuação de dados funciona de maneira reativa, ou seja, os dados são replicados somente quando for necessário. Nesse caso, a ação de evacuar os dados é acionada a partir da detecção a \textit{priori} de um possível ataque como, por exemplo, um alerta de terremoto. A reação seguinte consiste em determinar quais dados devem ser evacuados primeiro, essa seleção pode ocorrer de várias formas dependendo de como a prioridade será determinada. Pode-se notar que diferente da abordagem de redundância, que é proativa, no processo de evacuação, os dados replicados estão em sua versão mais atual, sendo assim, não haverá perdas significativas de informação \cite{Karakoc2016} \cite{YuanYang2016}.
 
Neste trabalho é proposta uma estratégia de evacuação de dados de DCs em zona de risco (i.e., sob alerta de ataque) para DCs em zona segura, ou seja, fora da área de alcance do ataque. Assim como no trabalho de \cite{Li2016}, assumimos que os DCs contam com um sistema de detecção de ataque, para que possam reagir logo em seguida a detecção, evacuando os dados para áreas seguras. Devido ao grande volume de dados concentrados em cada DC, é necessária uma política de priorização para decidir quais dados serão evacuados primeiro, criando-se uma fila de prioridades. O algoritmo de decisão, deve ser rápido o suficiente para migrar a maior quantidade possível de dados em menor tempo, tendo em vista o tempo crítico para a ocorrência do ataque. Dessa forma, duas políticas de evacuação de dados foram propostas: i) política baseada no acordo de nível de serviço (do inglês, \textit{Service Level Agreement} (SLA)) dos dados e ii) política baseada na ordem em que os dados chegam no DC, esta segue a ideia do algoritmo LIFO (do inglês, \textit{Last In, First Out}).

O restante do trabalho está organizado da seguinte forma: a Seção \ref{sec:trab-relacionados} apresenta os trabalhos relacionados a esta pesquisa e o estado da arte, a Seção \ref{sec:proposta} descreve a proposta. A Seção \ref{sec:simulacoes} apresenta os experimentos e os resultados obtidos e, por fim, a Seção \ref{sec:conclusao} aborda as considerações finais e discute possíveis trabalhos futuros.

%% file: 2-trab-relacionados.tex
\section{Trabalhos Relacionados} \label{sec:trab-relacionados}
A centralização de dados nos DCs, apesar de haver réplicas espalhadas geograficamente pelo globo, sofre uma ameaça eminente de ataques de qualquer natureza o que oferece alto risco para a perda de dados, principalmente àqueles que ainda não possuem réplicas ou que suas mudanças recentes não tenham sido sincronizadas em suas respectivas réplicas. Estes fatos têm atraído pesquisas na área, destacando-se os trabalhos de \cite{Ferdousi2015}, \cite{YuanYang2016}, \cite{Li2016} e \cite{Karakoc2016}.

No trabalho de \cite{Ferdousi2015}, os autores apresentaram uma heurística gulosa para a realização da evacuação de dados em DCs. A heurística proposta funciona por meio da seleção de caminhos de baixo custo, levando em consideração atraso de propagação, largura de banda e congestionamento, através de um modelo de rede \textit{unicast}. Em seguida, são escolhidos os conteúdos críticos e vulneráveis, para realizar a evacuação do maior número possível desses conteúdos dentro do tempo limite entre detecção e a ocorrência do ataque. 

No entanto, em \cite{Ferdousi2015} não foi especificado como é definida a prioridade dos conteúdos a serem evacuados. Os autores assumiram a prioridade como sendo um valor $\alpha$. Diferente da proposta desses autores, neste trabalho, a prioridade será dada de acordo com o SLA atribuído aos dados, uma vez que os conteúdos com um SLA elevado demandam um nível maior de disponibilidade e integridade. Dessa forma, o algoritmo proposto classifica os dados em ordem crescente, seguindo seus níveis de SLA atribuídos, para compor a fila de evacuação.

Em \cite{YuanYang2016} foi desenvolvida uma heurística para a recuperação, de forma eficiente, da conectividade da rede óptica entre os DCs. Os autores, modelaram a rede em um grafo direcionado, onde a heurística consiste na identificação e separação de nós que formam ciclo no grafo, a fim de torná-lo acíclico para evitar iterações infinitas na execução do algoritmo. Dessa forma, o algoritmo consegue determinar quais os DCs deve recuperar, baseando-se no grau de importância relativo atribuído ao DC.

Diferente do trabalho de \cite{YuanYang2016}, no trabalho proposto é abordada, exclusivamente, a evacuação dos dados visando manter a segurança e a disponibilidade dos mesmos. Em vez de tentar recuperar a conectividade dos DCs da rede, os dados de DCs em zona de risco são migrados para locais seguros, fora da área do ataque.    

Em \cite{Li2016} foi criado um esquema para evacuação de dados, em redes de nuvem ópticas, baseado em risco. Os dados são evacuados seguindo a mesma métrica de importância definida em \cite{Ferdousi2015} (i. e., $\alpha$). Os autores escolheram os melhores caminhos para a evacuação dos dados baseando-se em limitações como capacidade do enlace, tempo, capacidade de armazenamento do DC e o risco de falha de enlace. No trabalho proposto, o melhor caminho é determinado a partir de sua distância e, para isso, é utilizado o algoritmo de caminho mínimo de Dijkstra. 

No trabalho de \cite{Karakoc2016}, os autores propuseram um modelo de programação linear inteira para tratar o problema de migração de máquinas virtuais (MVs) durante um possível ataque cibernético a DCs. A técnica proposta conseguiu reduzir, consideravelmente, os efeitos do ataque ao migrar a maioria das MVs, com um custo mínimo relacionado aos recursos de rede e memória, para zonas seguras. No entanto, os autores não mencionaram qual o método utilizado para determinar o critério de escolha das MVs a serem migradas. Por outro lado, neste trabalho é abordada a migração de dados prioritários, sendo assim, uma dada MV também poderia ser compreendida como um conjunto de dados pelo algoritmo de evacuação proposto.

%% file: 4-proposta.tex
\section{Proposta} \label{sec:proposta} 

Os dados são evacuados com base em suas prioridades para os DCs mais próximos que encontram-se fora da área de risco do ataque. Portanto, considerando a criticidade de tempo para a ocorrência do desastre, os DCs localizados nas proximidades da zona afetada pelo desastre receberão os dados oriundos de DCs afetados. Durante o período de alerta de ataque é utilizada a largura de banda total disponível, no caminho entre o DC afetado e o DC de destino, para a transferência dos dados. De acordo com \cite{Ferdousi2015}, o atraso de transmissão é o mais significante e depende da taxa de dados do canal e do volume de dados trafegados.

A principal técnica empregada neste trabalho é a política de evacuação de dados baseada em suas respectivas importâncias para seus proprietários. A esta importância atribuímos o SLA por se tratar de um ambiente de nuvem \cite{mell2011nist}. \cite{Colman-Meixner2016} exemplifica que o serviço auto-adaptativo de elasticidade dos provedores de computação em nuvem pode adicionar mais recursos físicos para cumprir o acordo de nível serviço firmado com os clientes da nuvem. Entretanto, muitos provedores de serviços de nuvem podem ficar indisponíveis por longos períodos de tempo devido aos ataques. Portanto, os provedores de serviços, geralmente, devem pagar custos mais elevados devido ao descumprimento de acordos de nível de serviço com seus usuários finais \cite{Karakoc2016}.

Uma segunda técnica de evacuação de dados foi proposta e desenvolvida neste trabalho. Essa técnica baseia-se na ordem em que os dados chegam e são armazenados nos DCs, claramente, seguindo o algoritmo LIFO, ou seja, os dados recém chegados no DC têm maior prioridade em relação aos que chegaram primeiro (i.e., dados mais antigos), em caso de ocorrência de um ataque. Isso porque um determinado dado por ser mais recente, tem maior chance de ainda não possuir réplica na rede. Portanto, usando essa política os dados mais antigos serão evacuados, apenas se ainda houver tempo para a ocorrência do ataque e, somente depois que os dados mais novos já tiverem sido migrados para zonas seguras. 

\subsection{Serviços}

Em resumo, a solução proposta enfatiza três serviços principais que são descritos a seguir:

\begin{itemize}
\item {Armazenamento de dados -- o serviço de armazenamento nos DCs acontece a todo instante e mesmo durante o cenário de detecção de desastre. Dessa forma, a solução não interrompe o serviço de armazenamento nos DCs a menos que, o desastre ocorra e destrua sua infraestrutura. A solução adota o modelo de melhor esforço para atender os clientes da nuvem.}

\item {Clientes enviam dados aos DCs -- os clientes enviam dados constantemente para seus respectivos provedores de serviços em nuvem. Esses dados são armazenados pelos DCs enquanto estes estiverem em funcionamento e possuírem condições adequadas para realizar essa tarefa.}

\item {Identificação de ameaça e Evacuação rápida de dados -- os DCs contam com um sistema de monitoramento de ataques que, ao detectar uma possível ameaça, ativa o modo de operação em alerta e a evacuação dos dados entra em ação, seguindo a política de evacuação SLA (solução principal) ou LIFO (política desenvolvida para comparação). Portanto, o ``modo de alerta'' caracteriza o período em que o ataque foi detectado pelo DC afetado, mas ainda não ocorreu de fato.}
\end{itemize}

\subsection{Fases de Funcionamento}

A seguir são apresentadas as fases das duas políticas de evacuação de dados, propostas neste trabalho, juntamente com os serviços, mencionados anteriormente, em cada fase:

\begin{enumerate}
\item {\textbf{Operação normal} -- nesta fase, todos os DCs e infraestrutura da rede estão em funcionamento padrão, não apresentando nenhum problema. Os variados clientes enviam constantemente dados aos seus respectivos provedores de serviços em nuvem que recebem esses dados e os armazenam. Antes de haver o armazenamento, os dados que chegam aos DCs são classificados de acordo com a política adotada, que pode ser a estratégia SLA ou LIFO.}

\item {\textbf{Detecção do ataque} -- assume-se que os DCs possuem uma série de sensores para que possam detectar com o máximo de antecedência os mais variados tipos de ataques que podem ocorrer. A etapa de detecção do ataque muda o funcionamento dos DCs que passam do estado de \textit{operação normal} para o \textit{modo de alerta}. É nesse período que os DCs afetados calculam o tempo para a ocorrência do ataque e iniciam a evacuação de seus dados para àqueles DCs que encontram-se livres das ameaças do ataque. Todo o tempo restante disponível é utilizado para migrar os dados, visto que, como nos ambientes reais, nos experimentos realizados sempre há uma grande quantidade de dados nos DCs alvos de ataques.}

\item {\textbf{Ocorrência do ataque} -- após o ataque ser detectado e ter seu tempo limite para a ocorrência esgotado, toda a infraestrutura dos DCs alvos são destruídas simultaneamente. Nos experimentos realizados, é feita a análise dos dados oriundos dos DCs atacados, que conseguiram ser salvos nos DCs de destino durante o ataque.}
\end{enumerate}

Em ambas as políticas de evacuação de dados propostas neste trabalho, são considerados para serem migrados os dados do DC em risco classificado em \textit{modo de alerta}. Portanto, considerar a política LIFO significa priorizar sempre os dados que estão chegando exatamente no período de operação \textit{modo de alerta}. Por outro lado, a política SLA considera os dados que chegam no período \textit{modo de alerta} apenas se estes possuírem um SLA superior em relação aos dados que já estavam armazenados no DC sob ataque.

\begin{figure}[htbp]
\centering
\includegraphics[width=.8\textwidth]{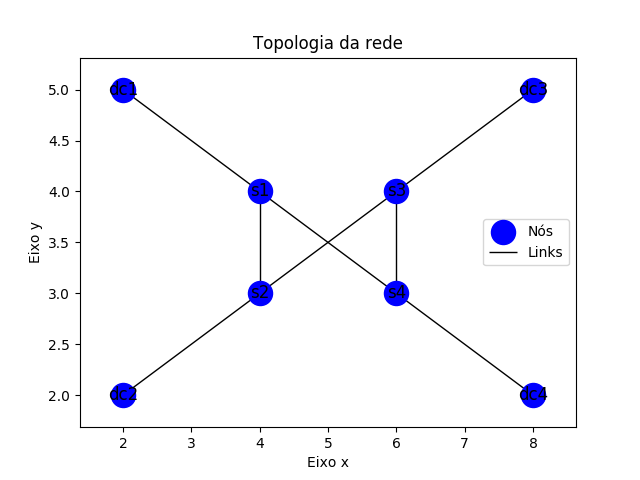}
\caption{Topologia da proposta empregada nos experimentos.}
\label{fig:topo-proposta}
\end{figure}

Nos experimentos realizados foram considerados apenas quatro DCs e quatro \textit{switches} interligados por fibras ópticas, embora a solução desenvolvida permita instanciar uma maior quantidade de DCs, optamos por simplificar a topologia visto que a utilizada é suficiente para realizar os experimentos propostos neste estudo. A topologia da rede está representada na Figura \ref{fig:topo-proposta}.

\begin{figure}[htbp]
\centering
\subfigure[DC1 sob ataque. \label{fig:dc1-ataque}]{
\includegraphics[width=.48\textwidth]{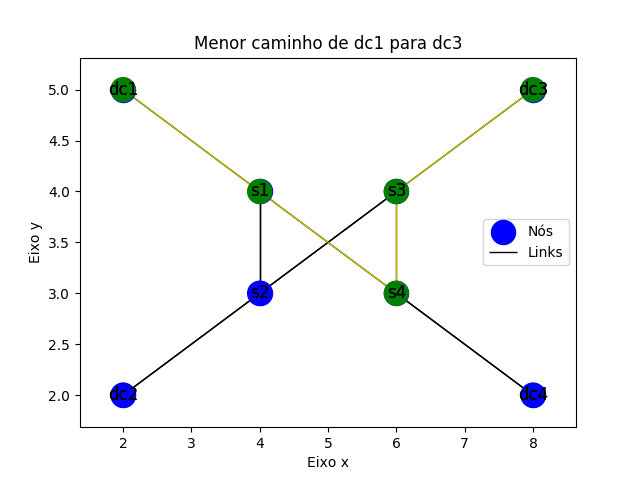}
} 
\subfigure[DC2 sob ataque. \label{fig:dc2-ataque}]{
\includegraphics[width=.48\textwidth]{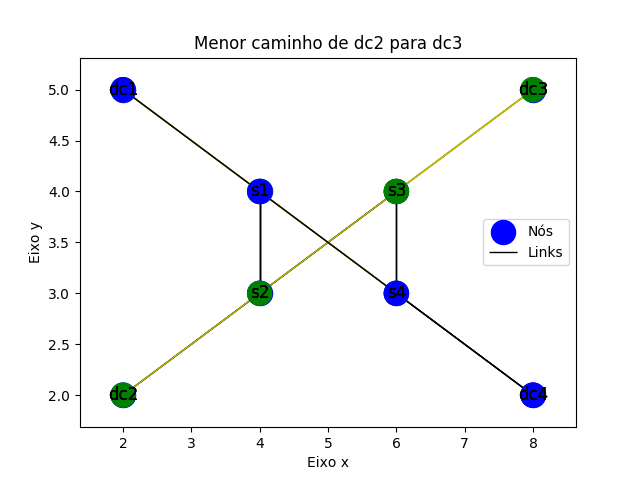}
} 
\caption{Cálculo do menor caminho de DCs sob ataque para DC em zona segura.}\label{fig:fpath}
\end{figure}

Ao detectar um ataque, o DC calcula o tempo para a ocorrência do mesmo, calcula o menor caminho para um determinado DC que esteja fora da área do alcance do ataque e inicia o processo de evacuação dos dados seguindo uma das duas políticas de priorização já apresentadas. A Figura \ref{fig:fpath} ilustra esse processo, observe que o menor caminho está representado na cor verde. Na Figura \ref{fig:dc1-ataque} é possível notar que o DC1 (sob ataque) obteve como menor caminho o DC3 (fora da área do ataque). O mesmo pode ser observado na Figura \ref{fig:dc2-ataque}, na qual retrata o cálculo de menor caminho do DC2 com destino ao DC3.  

A lógica da priorização dos dados, que chegam nos DCs para serem armazenados, está representada no Algoritmo \ref{alg:alg1}. Cada DC executa esse algoritmo que é responsável pela classificação dos dados que chegam, oriundos dos seus clientes, antes de armazená-los em suas \textit{storages}. Na linha 3, temos uma repetição que será executada enquanto a simulação ocorrer. Na linha 4, são lidos os dados recém chegados no DC. Na linha 5, é feita uma comparação para saber se a política de evacuação utilizada é a SLA. Na linha 7, é calculada a prioridade dos dados, para isso foi utilizada a função \texttt{PriorityItem()} da biblioteca \texttt{Python simpy}\footnote{\url{https://simpy.readthedocs.io/en/latest/}}, que recebe como argumento a política de prioridade e os dados a serem classificados. Na linha 8, os dados são de fato armazenados após passarem pela classificação de suas prioridades.

Na linha 9 uma segunda comparação é feita, para saber se a política utilizada é a LIFO. Na linha 11, novamente, a função \texttt{PriorityItem()} é utilizada, dessa vez, é passada a política LIFO e os dados, e na linha 12, os dados são armazenados após serem classificados. Por fim, a linha 13, representa a condição em que a política utilizada não é nenhuma das duas abordadas neste estudo, então, é emitido um alerta na tela (linha 14), informando tal situação e encerrando o algoritmo (linha 15). 

\begin{algorithm}[htbp]
\SetAlgoLined
\DontPrintSemicolon
\SetKwFunction{FReq}{server}
\SetKwProg{Fn}{Function}{:}{end}
\Fn{\FReq{self}}{  	
	$lifo\_priority = 0$ \tcp*{Inicia prioridade LIFO em zero.}
	\While{True}{
    	\tcc{Ler dados recém armazenados.}
    	$msg = self.store.get()$ \;
        \uIf{self.evacuation\_policy == self.sla\_policy}{
        	$sla\_priority = round(random.uniform(90, 99), 0)$ \;
    		\tcc{Classifica SLA dos dados.}
    		$data\_priority = \Call{PriorityItem}{((sla\_priority) * -1), msg}$ \;
            \tcc{Armazena dados na store classificados.}
       		$self.data\_priority.put(data\_priority)$ \;
 		}
  		\uElseIf{self.evacuation\_policy == self.lifo\_policy}{
            $lifo\_priority += 1$ \;
            \tcc{Classifica ordem de chegada dos dados.}
            $data\_priority = \Call{PriorityItem}{((lifo\_priority) * -1), msg}$ \;
            \tcc{Armazena dados na store classificados.}
            $self.data\_priority.put(data\_priority)$ \;
 		}
  		\Else{
    		$imprima (Aviso: evacuation\_policy deve ser ``sla\_policy'' ou ``lifo\_policy'')$ \;
            \Return $0$ \;
  		}
 	}
}
\caption{Política de evacuação baseada no SLA e LIFO.}
\label{alg:alg1}
\end{algorithm}

%% file: 5-experimentos.tex
\section{Simulações e análise dos resultados} \label{sec:simulacoes} 

Os experimentos realizados neste trabalho seguiram os conceitos abordados por \cite{jain1990art}. Portanto, utilizou-se uma abordagem sistemática em todas as etapas desta pesquisa. Como parte dessa abordagem, a Tabela \ref{table1} destaca os parâmetros utilizados nas simulações.

\begin{table}[htbp]
\centering
\caption{Parâmetros da Simulação.}
\label{table1}
\begin{tabular}{l l} 
\toprule
\textbf{Parâmetro} & \textbf{Valor} \\ 
\midrule
Simulador de redes & Próprio, baseado no \textit{simpy} \\ 
Nível do SLA & Randômico, variando de 90,0\% à 99,0\% \\ 
Tempo para ocorrer o ataque & 1:30, 2:00 e 2:30 minutos após o início da simulação \\ 
Estratégias de evacuação & Baseada no SLA e baseada no algoritmo LIFO \\ 
Capacidade dos enlaces & 1, 5 e 10 Gbps \\ 
Taxa de geração de dados & Baseada na distribuição \textit{Poisson} \\
\bottomrule
\end{tabular}
\end{table}

Inicialmente foi realizado um projeto fatorial $2^k$ para aferir o impacto que cada fator exerce sobre os resultados das simulações. Nessa etapa inicial, $k$ recebeu o valor $3$, ou seja, a combinação de três fatores e dois níveis, no qual, cada nível assumiu os valores mínimo e máximo. A Tabela \ref{table2} ilustra a importância dos fatores, representada em porcentagem na coluna 3 (Impacto), bem como, o valor do impacto causado aos resultados pela interação entre os fatores. Os valores foram obtidos em \textit{megabytes} e transformados em porcentagem.

\begin{table}[htbp]
\centering
\caption{Fatores do Projeto Fatorial.}
\label{table2}
\begin{tabular}{l l l l} 
\toprule
\textbf{ID} & \textbf{Fator} & \textbf{Impacto} & \textbf{Valor (mín. e máx.)} \\ 
\midrule
\textbf{A} & Tempo de evacuação & 9,90\% & 10 e 20 segundos \\ 
\textbf{B} & Largura de banda & 32,82\% & 1 e 10 Gbps \\
\textbf{C} & Quantidade de clientes ativos & 26,81\% & 20 e 40 clientes \\
\bottomrule
\toprule
\multicolumn{4}{c}{\textbf{Impacto representado pelas interações entre os fatores}} \\ 
\midrule
\textbf{AB} = 1,58\% & \textbf{AC} = 1,05\% & \textbf{BC} = 26,81\% & \textbf{ABC} = 1,05\% \\
\bottomrule
\end{tabular}
\end{table}

Para o estudo inicial foi utilizada como métrica a quantidade de dados evacuados durante um ataque, devido ao fato de que a tarefa de salvar dados representa um dos principais objetivos em uma ocorrência de ataque aos DCs. Além disso, nesta etapa, os experimentos foram realizados utilizando a estratégia de evacuação de dados baseada no SLA. 

Após o estudo fatorial dos fatores, foi realizado um estudo mais aprofundado utilizando os dois fatores de maiores relevâncias para a pesquisa, como se pode observar na Tabela \ref{table3}. Agora, utilizando três níveis para cada fator. 

\begin{table}[htbp]
\centering
\caption{Fatores da Simulação.}
\label{table3}
\begin{tabular}{l l l} 
\toprule
\textbf{Fator} & \textbf{Impacto} & \textbf{Valor} \\ 
\midrule
Largura de banda & 32,82\% & 1, 5 e 10 Gbps \\
Quantidade de clientes ativos & 26,81\% & 20, 30 e 40 clientes \\
\bottomrule
\end{tabular}
\end{table}

Além de empregar os três valores de ambos os dois fatores, foram estabelecidas novas métricas, listadas abaixo, a fim de analisar o desempenho da proposta com uma riqueza maior de detalhes:

\begin{itemize}
\item Taxa de dados evacuados (\%) -- relação da média total de dados salvos, pela quantidade de dados armazenados em um DC sob ataque. Utilizada para medir a eficácia da solução.

\item Quantidade de dados evacuados por grau de prioridade (GB) -- média total de dados salvos de acordo com seu grau de importância. Utilizada para medir a eficiência da solução.

\item Tempo de migração de dados (ms) -- média de tempo dada em milissegundos, do atraso de cada pacote de dados, a partir da detecção do ataque para o dado ser migrado para DC localizado em zona segura.
\end{itemize}

\subsection{Análise de Desempenho}

Para ilustrar as peculiaridades das estratégias SLA e LIFO, foram realizados experimentos com ambas as estratégias. Em todos os experimentos os DCs, DC1 e DC2, sofreram ataques simultaneamente e migraram seus dados usando as duas políticas propostas. 

Os dois fatores e três níveis da Tabela \ref{table3} geraram um total de nove cenários por DC em ataque, cada cenário foi executado seis vezes, o que gerou um total de 54 simulações usando a política SLA e o mesmo valor usando a política LIFO, contabilizando 108 execuções no geral. Vale ressaltar que além das análises por estratégia, foram realizadas análises individualmente em cada DC sob ataque. A média foi calculada com um nível de confiança de 95\% para todas as métricas, utilizando a Tabela de Distribuição \textit{t-Student} devido a baixa quantidade de replicações dos experimentos.

\begin{figure}[htbp]
\centering
\includegraphics[width=.8\textwidth]{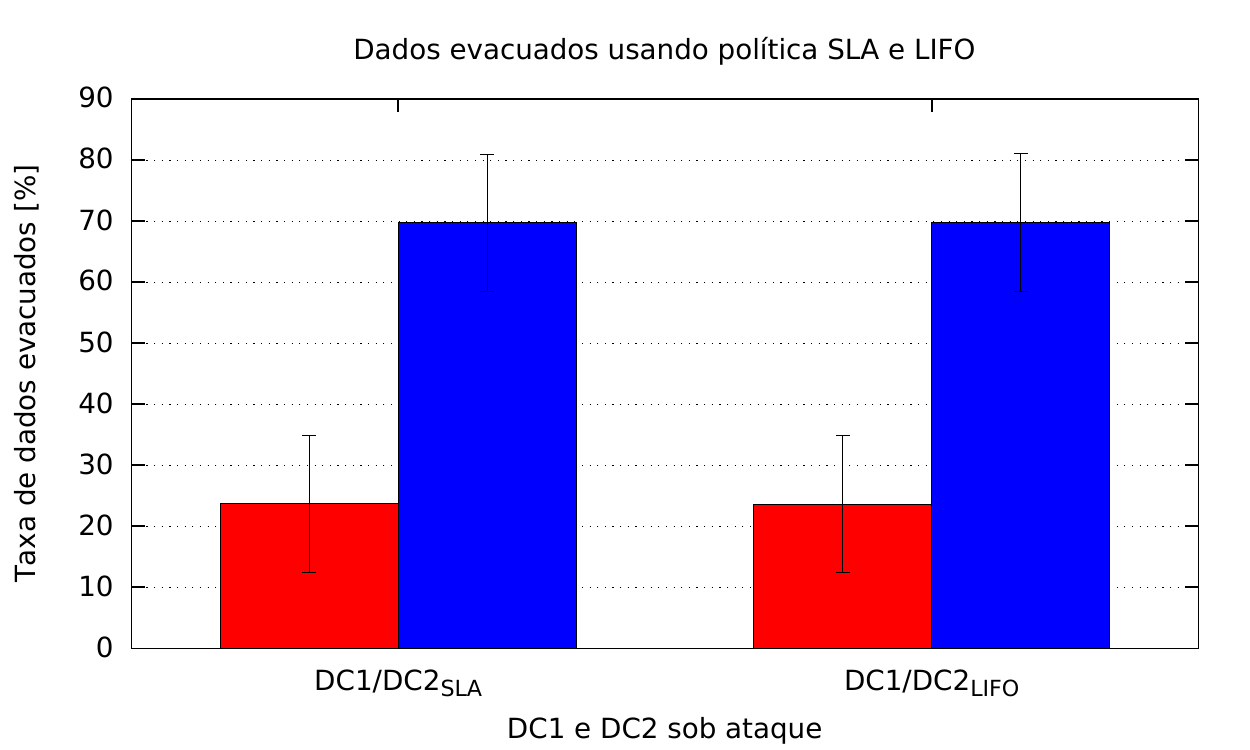}
\caption{Taxa de dados evacuados por DC com política SLA e LIFO.}
\label{fig:avg-data}
\end{figure}

A porcentagem dos dados salvos, dos DCs DC1 e DC2, está representada na Figura \ref{fig:avg-data}. É possível notar que ambas as políticas de evacuação propostas apresentam um desempenho muito semelhante em relação a porcentagem dos dados salvos. No entanto, o DC2 apresentou um desempenho superior ao DC1 nas duas políticas, no qual, obteve pouco mais que o dobro de dados evacuados se comparado ao DC1. Utilizando a política SLA cerca de 23\% e 70\% dos dados armazenados no DC1 e DC2 foram salvos, respectivamente. Os mesmos valores foram obtidos quando utilizada a política LIFO.

\begin{figure}[htbp]
\centering
\includegraphics[width=.8\textwidth]{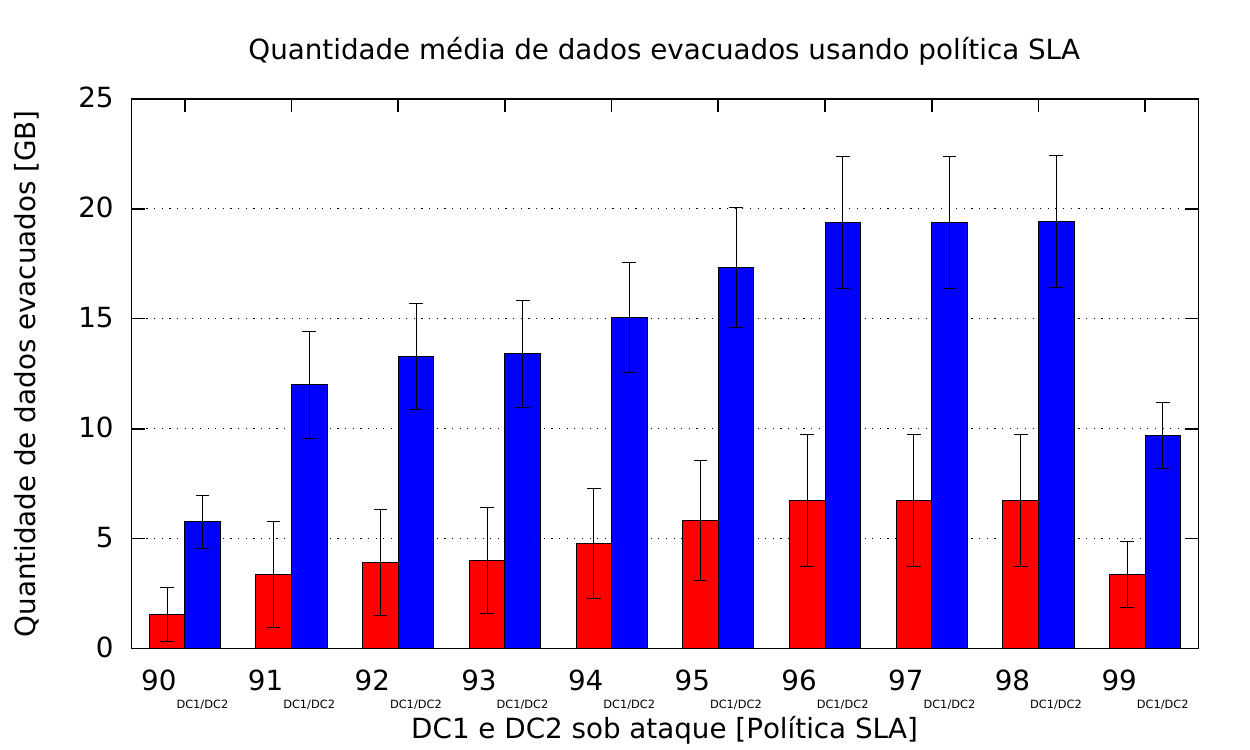}
\caption{Quantidade de dados evacuados por DC com política SLA.}
\label{fig:avg-priority-sla}
\end{figure}

Além da análise da taxa de dados salvos, foi investigada a prioridade desses dados. A Figura \ref{fig:avg-priority-sla} apresenta a prioridade dos dados salvos oriundos do DC1 e DC2, segundo o SLA dos mesmos, em uma escala que varia de 90,0\% (menor prioridade) à 99,0\% (maior prioridade) de SLA. O DC2 obteve um desempenho superior ao DC1 em todos os cenários. É importante ressaltar que a carga de trabalho submetida aos algoritmos propostos, se baseia na distribuição \textit{Poisson} com média (0,1/100), portanto, uma carga considerada elevada. Apesar disso, a quantidade de dados gerados com um nível correspondente a 99,0\% de SLA foi inferior aos demais níveis, o que ficou comprovado na Figura \ref{fig:avg-priority-sla}. O algoritmo migrou todos os dados correspondentes a 99,0\% primeiro, seguindo a ordem decrescente na escala de prioridades. Esse comportamento, cumpre exatamente ao objetivo dessa estratégia, salvar primeiro os dados críticos (i.e., com SLA elevado).

\begin{figure}[htbp]
\centering
\includegraphics[width=.8\textwidth]{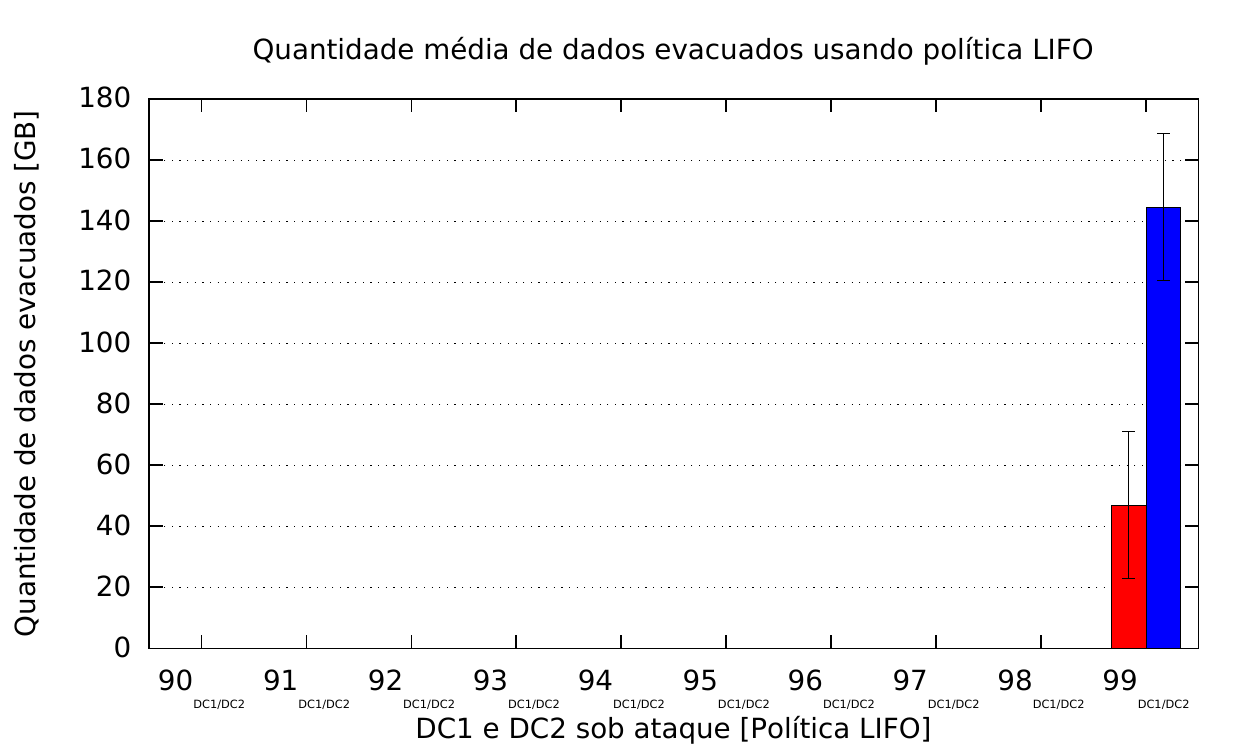}
\caption{Quantidade de dados evacuados por DC com política LIFO.}
\label{fig:avg-priority-lifo}
\end{figure}

A Figura \ref{fig:avg-priority-lifo} apresenta a prioridade dos dados salvos oriundos do DC1 e DC2, segundo a ordem que os mesmos chegaram nos DCs sob ataque. Para facilitar a análise e comparação dos resultados obtidos na política LIFO com a política SLA, todos os dados migrados pela estratégia LIFO foram colocados no nível 99,0\% da escala de prioridades, já que esta estratégia migra os dados sem considerar o nível de SLA, ou seja, como se os dados pertencessem a um mesmo nível. Novamente, o DC2 alcançou um desempenho superior, migrando cerca de 142 GB de dados enquanto o DC1 migrou cerca de 43 GB.

\begin{figure}[htbp]
\centering
\includegraphics[width=.8\textwidth]{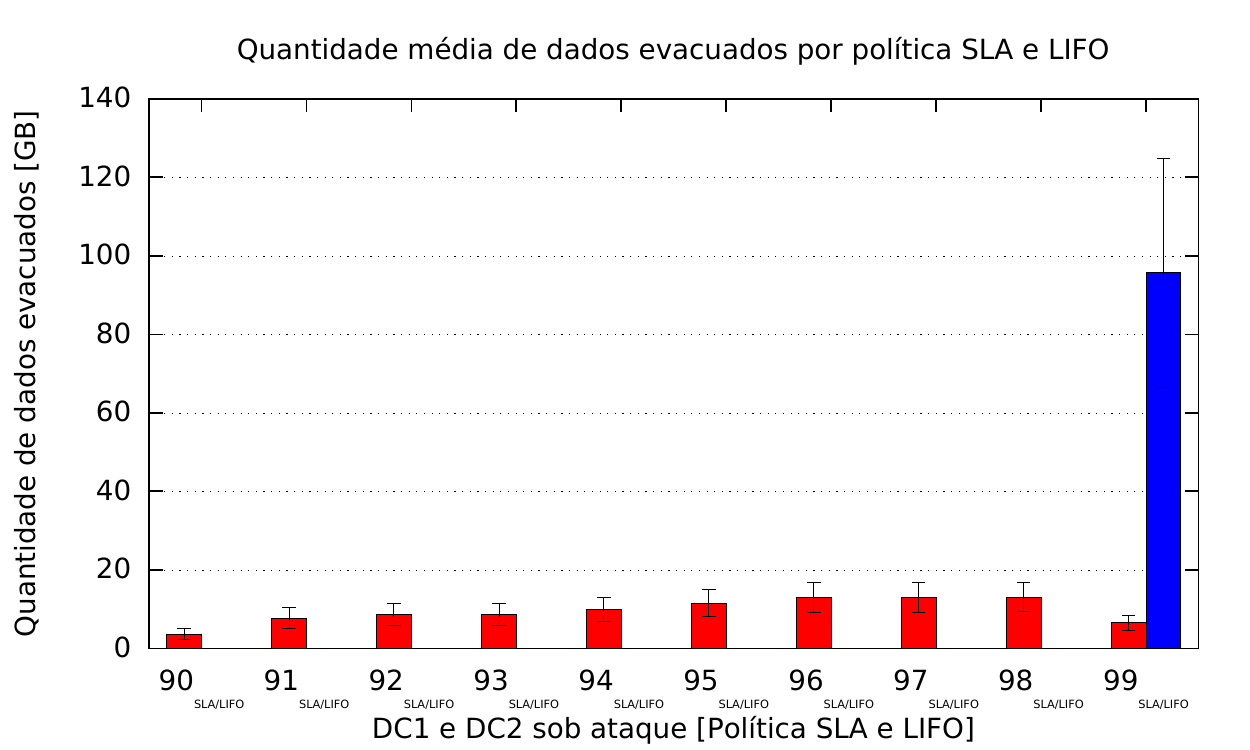}
\caption{Quantidade total de dados evacuados com as políticas SLA e LIFO.}
\label{fig:avg-priority}
\end{figure}

A quantidade total, em \textit{gigabytes}, de dados evacuados por política está representada na Figura \ref{fig:avg-priority}. A política SLA distribui os dados de acordo com a escala de prioridades, já a política LIFO não segue o nível de SLA e considera, portanto, todos os dados como pertencentes ao mesmo SLA (99,0\%). A soma dos dados distribuídos nos níveis compõem a quantidade de dados da política SLA. A política LIFO migrou cerca de 98 GB de dados, oriundos dos DCs sob ataque, para locais seguros.

\begin{figure}[htbp]
\centering
\includegraphics[width=.8\textwidth]{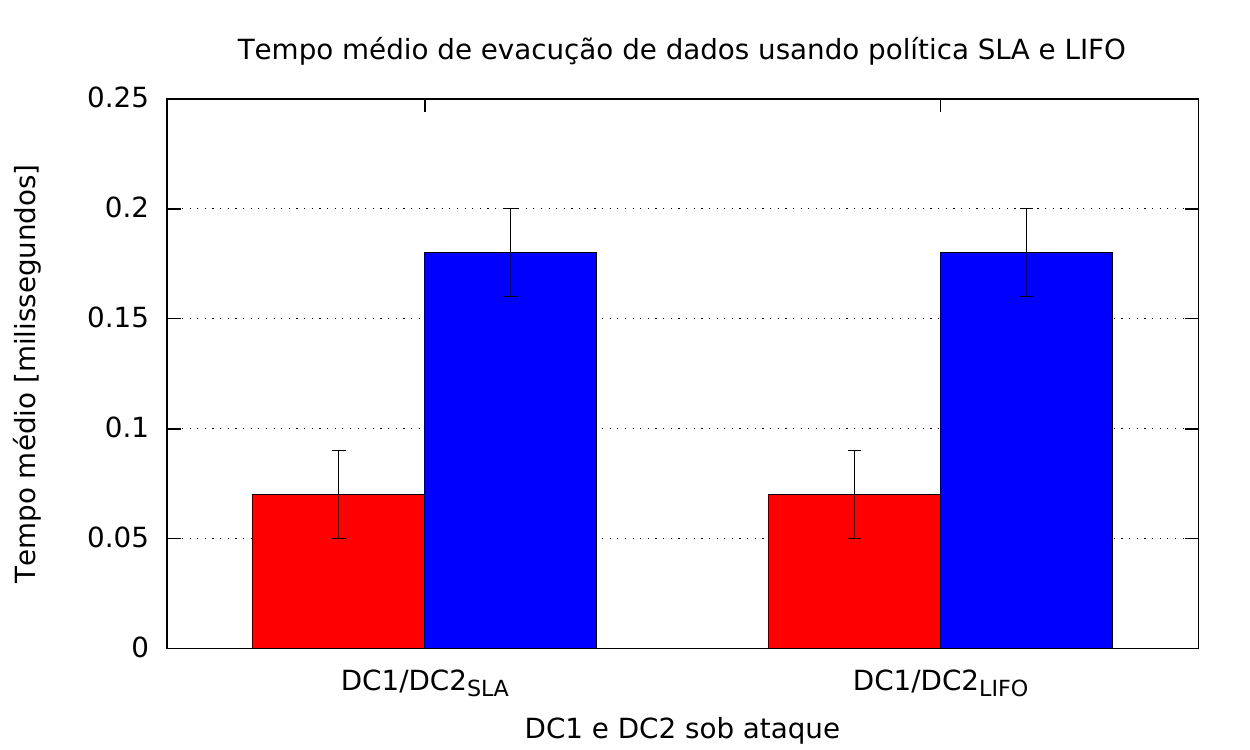}
\caption{Tempo médio de evacuação de dados com política SLA e LIFO.}
\label{fig:avg-time}
\end{figure}

Por fim, foi avaliado o tempo médio, em milissegundos, que os pacotes de dados levaram para serem migrados para locais seguros. Os pacotes possuíam tamanho equivalente a 1500 \textit{bytes}. A análise se deu no DC1 e DC2, ambos, utilizando as duas políticas propostas. É possível observar na Figura \ref{fig:avg-time}, que as políticas SLA e LIFO apresentaram uma média de tempo similar nos DCs em ataque. O DC2 obteve um tempo maior nas duas políticas, o que é justificado pelo fato desse DC ter migrado uma quantidade maior de dados. Isso ficou evidente na Figura \ref{fig:avg-data}, mostrada anteriormente.  

%% file: 6-conclusao-e-trab-fut.tex
\section{Conclusão e Trabalhos Futuros} \label{sec:conclusao}
Os DCs se caracterizam por possuir grandes quantidades de dados, esses ambientes concentram dados de diferentes clientes, os quais firmam acordos com os provedores de serviços em nuvem. Esses acordos são chamados de SLA, possuem valor jurídico e representam o nível de disponibilidade e proteção aplicado aos dados armazenados. Quanto maior o SLA maior é o custo financeiro para os clientes e DCs, por isso, somente os dados mais importantes, geralmente, possuem o SLA elevado. Os DCs estão sujeitos a ataques de vários tipos, naturais ou causados pelo homem. Os ataques são difíceis de prever, podem acontecer rapidamente e podem ter alto poder de destruição de dados.   

Nesse contexto, este trabalho apresentou duas estratégias de evacuação de dados para DCs em situação de ataque. A principal estratégia é baseada no SLA dos dados, em que migra, primeiro, os dados de maior SLA em caso de ataque. A segunda estratégia é baseada no algoritmo LIFO e evacua dados de acordo com a ordem que foram armazenados no DC. Ambas as estratégias apresentaram praticamente os mesmos resultados quanto a quantidade de dados evacuados e tempo para evacuação. No entanto, a política SLA distribui os dados sobre uma escala de prioridades baseada no SLA dos dados, migrando primeiro os dados mais a direita da escala (SLA elevado, 99,0\%), enquanto a política LIFO classifica os dados numa mesma prioridade da escala, o que pode resultar na priorização de dados com SLA menor para serem evacuados, prejudicando a migração dos dados com SLA elevado.

Para trabalhos futuros, pretendemos expandir a topologia da rede para o cenário ficar mais realístico e aumentar a quantidade de replicação das simulações.